\documentclass[prd,showpacs,showkeys,floatfix,
               12pt,preprint, tightenlines,
               fleqn]{revtex4}

\usepackage{amsmath}
\usepackage{amsfonts}
\usepackage{latexsym}
\usepackage{graphicx}

\newcommand{\lplanck}{l_{\rm Planck}}
\newcommand{\lgamma}{l_\gamma}

\newcommand{\lwormhole}{l_\mathrm{wormhole}} 
\newcommand{\lseparation}{l_\mathrm{separation}}

\newcommand{\beq}{\begin{equation}}
\newcommand{\eeq}{\end{equation}}
\newcommand{\beqa}{\begin{eqnarray}}
\newcommand{\eeqa}{\end{eqnarray}}

\newcommand{\bR}{\mathbb{R}}

\newcommand{\half}{{\textstyle \frac{1}{2}}}

\newcommand{\dx}{\!{\rm d}^4x\,\,}

\begin{document}
\noindent Phys. Rev. D {\bf 72}, 017901 (2005)
\hfill hep-ph/0506071, KA--TP--08--2005
\vspace*{2\baselineskip}
\title{Photon-propagation model with random background field:
Length scales and Cherenkov limits}

\author{Frans R.\ Klinkhamer}
\email{frans.klinkhamer@physik.uni-karlsruhe.de}
\author{Christian Rupp}
\email{cr@particle.uni-karlsruhe.de}
\affiliation{Institut f\"ur Theoretische Physik,\\Universit\"at Karlsruhe (TH),\\
             76128 Karlsruhe, Germany\\ \\}

\begin{abstract}
We present improved experimental bounds on typical length scales of a
photon-propagation model with a frozen (time-independent) random background field,
which could result from anomalous effects of a static, multiply connected
spacetime foam.
\end{abstract}


\maketitle

\section{Introduction}
\label{sec:intro}

In a previous article \cite{KlinkhamerRupp}, we have proposed a simple
photon-propagation model to describe the potential
effects of a static spacetime foam composed of identical,
randomly-distributed defects (e.g., microscopic wormholes)
embedded in Minkowski spacetime. For this particular model,
a modified photon dispersion law was derived
in the long-wavelength limit,
\begin{equation}
\omega^2 \sim (1-A^2\, \gamma_1) \, c^2 k^2 -
A^2\, \lgamma^2\,c^2\, k^4\,, \label{foamdisp}
\end{equation}
where $k \equiv |\vec k|$ is the photon wave number and
$\omega$ the frequency, $A$ the amplitude of the frozen (time-independent)
random background field $g_1(\vec x\,)$,
$\lgamma$ a characteristic length scale of $g_1(\vec x\,)$,
$\gamma_1$ a nonnegative dimensionless
coefficient, and $c$ a fundamental constant tracing back to the  Minkowski
line element (see also below). An upper bound
$\lgamma < 1.6\times 10^{-22}\,{\rm cm}$,
for $A=\alpha\approx 1/137$,
was then obtained
from observations of a particular TeV flare in an active galactic nucleus.

In this Brief Report, we use recent results on ultra-high-energy cosmic rays
\cite{ColemanGlashow,GagnonMoore}  to improve our previous bound on $\lgamma$.
In addition, we give a careful discussion of the possible relation between
the photonic length scale $l_\gamma$ and the characteristic length scales of
the microscopic spacetime structure.

In the following, we will use standard natural units with
$\hbar$ $=$ $c$ $=$ $1$, except when stated otherwise.
For the physical situation discussed
in the next section, the operational definition of the velocity $c$
is the maximum attainable velocity of the proton. (Further discussions
on Lorentz noninvariance can be found in, e.g.,
Refs.~\cite{ColemanGlashow,GagnonMoore,ColladayKostelecky,Bertolami-etal}
and references therein.)

\section{Photon propagation}
\label{sec:photon-propagation}
Assuming a modified photon dispersion law
with a negative dimensionful coefficient $K_\mathrm{1\,neg}$,
\beq
E_\gamma \sim k + K_\mathrm{1\,neg} \,k^3 \,,
\eeq
and an unchanged (ultrarelativistic) proton dispersion law,
\beq
E_p \sim k \,,
\label{protondisp}
\eeq
the Cherenkov-like proton process $p \to p + \gamma$
becomes kinematically allowed \cite{ColemanGlashow}.
From observations of ultra-high-energy cosmic rays,
Gagnon and Moore \cite{GagnonMoore} obtain the following bound:
\begin{equation}
0 \leq -K_\mathrm{1\,neg} < (4 \times 10^{22} \, {\rm GeV}) ^{-2}\,.
\label{GMbound}
\end{equation}

There is also a bound on the difference between the maximum attainable
velocities of particles with spin $1$ and spin $1/2$. For a modified photon
dispersion law
\begin{equation}
E_\gamma = c\, (1+\epsilon)\, k
\end{equation}
and an unmodified fermion dispersion law (\ref{protondisp}), the authors of
Ref.~\cite{GagnonMoore} obtain the bound
\begin{equation}
|\epsilon| < 1.6 \times 10^{-23}\,.\label{GMbound2}
\end{equation}

We now turn to a simple photon-propagation model \cite{KlinkhamerRupp}
with a fixed random background field $g_1(x)$ and an action given by
\begin{align} 
S_\mathrm{\, photon} = & -\textstyle{\frac{1}{4}}\,\int_{\bR^4}\, \dx
\Bigl(F_{\mu\nu}(x) F_{\kappa\lambda}(x)\, \eta^{\kappa\mu} \eta^{\lambda\nu}
+ \, g_1(x)\,  F_{\kappa\lambda}(x)  \widetilde F^{\kappa\lambda}(x)\Bigr) \,,
\label{Photonmodel}\end{align}
in terms of the standard Maxwell field strength tensor
$F_{\mu\nu}\equiv \partial_\mu A_\nu -\partial_\nu A_\mu$, the
dual tensor $\widetilde F^{\kappa\lambda}\equiv
\half\;\epsilon^{\kappa\lambda\mu\nu}\, F_{\mu\nu}$,
and the inverse Minkowski metric $\eta^{\mu\nu}$.
The random background field $g_1(x)$ is assumed to be time-independent,
\beq
g_1(x^0,x^1,x^2,x^3)= g_1(x^1,x^2,x^3) \equiv g_1( \vec x\,) \,,
\label{g1}
\eeq
and to fluctuate around a value zero with amplitude $A$;
see Sec.~IV of Ref.~\cite{KlinkhamerRupp} for further properties.
The random background field $g_1(\vec x)$ in Eq.~(\ref{Photonmodel})
can be seen to act as a variable coupling constant,
with spacetime taken to be perfectly smooth
(manifold $\mathrm{M} = \bR^4$ and metric $g_{\mu\nu}(x) = \eta_{\mu\nu})$.

For the photon-propagation model (\ref{Photonmodel}),
we have calculated in Sec.~V of Ref.~\cite{KlinkhamerRupp}
the dispersion law (\ref{foamdisp}), with $\lgamma$ and $\gamma_1$ determined
in terms of the autocorrelation function of $g_1( \vec x\,)$,
\begin{equation}
\lgamma = \lgamma  \,\!\left[g_1 \right]  \,, \quad
\gamma_1 = \gamma_1\,\!\left[g_1 \right]\,.
\label{lfoamgamma1fromg1}
\end{equation}
The dispersion law (\ref{foamdisp}) gives then the following photon energy:
\begin{equation}
E_\gamma \sim k \left( 1- \half\, A^2 \gamma_1 -
                          \half\, A^2 \, \lgamma^2 \, k^2 \right)\,,
\label{dispersionlaw}
\end{equation}
for parametrically small amplitude $A$
or for $\gamma_1$ and $\lgamma^2 \, k^2$ much less than unity.

 From the experimental bounds (\ref{GMbound}), (\ref{GMbound2}) and
the relation (\ref{dispersionlaw}), we obtain
\begin{subequations}\label{lfoamgamma1bounds}
\begin{align} 
\lgamma &<
\left(2 \times 10^{20} \,{\rm GeV}\right)^{-1}\;\left(\alpha\; A^{-1}\,\right)  \,
\approx
\left(1.0 \times 10^{-34}\, {\rm  cm}\right)\;\left(1/137\;A^{-1}\,\right)\,,
\label{lfoambound}\\[2mm]
\gamma_1 &< \left(6\times 10^{-19}\,\right)\;\left(\alpha/A\right)^2 \,,
\label{gamma1bound}
\end{align}
\end{subequations}
where $\alpha\approx 1/137$ is the fine-structure constant (the possible
relation $A\sim\alpha$ will be discussed in the next section).
Note that the bound (\ref{lfoambound}) is $12$ orders of magnitude
better than the one given in Sec.~VI of Ref.~\cite{KlinkhamerRupp},
where $\lgamma$ was called $l_{\rm  foam}$.
This  bound on $\lgamma$  for $A \sim \alpha$ is, in fact,
of the order of the Planck length,
\beq
\lplanck \equiv
\sqrt{G\,\hbar/c^3} \approx 1.6 \times 10^{-33}\,\,{\rm cm}\,,
\label{lplanck}
\eeq
which may determine the fine-scale structure of
spacetime itself \cite{Wheeler}.

\section{Spacetime structure}
\label{sec:spacetime-}

In order to connect the photon parameters
$\gamma_1\left[g_1 \right]$ and $\lgamma\left[g_1 \right]$
derived from the effective action (\ref{Photonmodel})
to the microscopic structure of spacetime,
we introduce the following definitions:
\begin{subequations}\label{lfoamgamma1generaldef}
\begin{align}
\lgamma &\equiv \lwormhole\;\left(\lwormhole /\lseparation\right)^{3/2}\,,
\label{lfoamgeneraldef}\\[2mm]
\gamma_1 &\equiv \left(\lwormhole/\lseparation\right)^3\,.
\label{gamma1generaldef}
\end{align}
\end{subequations}
These definitions are motivated by a very simple spacetime
model \cite{KlinkhamerRupp} consisting of static,
randomly-distributed wormholes \cite{Wheeler} embedded in Minkowski
spacetime. This toy model has, by definition, a preferred frame of reference.
The length $\lwormhole$ would then correspond to
an appropriate characteristic dimension
of an \emph{individual} wormhole (e.g., the average width of the two mouths or
the long distance between the centers of the mouths, where both lengths are
measured in the Minkowski part of spacetime and the short
distance through the wormhole throat is assumed to be zero).
The length $\lseparation$ would  correspond to
the average separation between \emph{different}
wormholes (the wormhole density is $n_\mathrm{wormholes}=\lseparation^{-3}$).

The anomaly calculation reported in the Appendix of
Ref.~\cite{KlinkhamerRupp}, specialized to the case $l_h=\delta$
and with notations $(l_{\rm  foam},\, d,\, a)$ for
$(\lgamma,\, \lwormhole,\, \lseparation)$ here, gave $A=\alpha$
in the effective action (\ref{Photonmodel})
and extra factors
$0.18$ and $0.15$ on the right-hand sides of
Eqs.~(\ref{lfoamgeneraldef}) and
(\ref{gamma1generaldef}), respectively. This calculation
was, however, based on several simplifying assumptions and
is, therefore, not absolutely rigorous. The two most important results
would be that there are no extremely small or large factors
on the right-hand sides of Eqs.(\ref{lfoamgamma1generaldef}ab)
and that the effective amplitude $A$ is of order $\alpha$.
The physical interpretation of the quantities
$\lwormhole$ and $\lseparation$,
defined mathematically by Eqs.~(\ref{lfoamgamma1generaldef}ab), would be
that they emerge directly from the underlying spacetime structure.
Indeed, a successful calculation would relate
the ``randomness'' of the couplings
$g_1( \vec x\,)$ in the effective action (\ref{Photonmodel})
to the (as of yet, unknown) microscopic structure of spacetime.
(An entirely different origin for the variable couplings
$g_1(x)$ of a $F_{\kappa\lambda}\,\widetilde F^{\kappa\lambda}$ term
in the effective action is, of course, not excluded;
see, e.g., Ref.~\cite{Bertolami-etal} and references therein.)

\begin{figure}[t]
\begin{center}
\includegraphics[height=5.8cm]{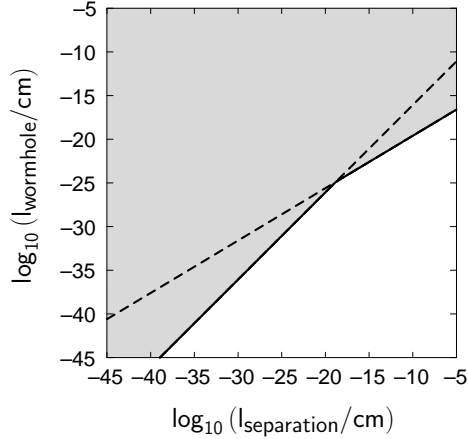}
\end{center}
\caption{Excluded  region [shaded region above the solid curves]
for photon-propagation length scales (\ref{lfoamgamma1generaldef}ab)
from cosmic-ray bounds (\ref{lfoamgamma1bounds}ab) for $A=\alpha$.
These length scales can perhaps be interpreted as corresponding to a
static spacetime model which consists of
identical, randomly-distributed wormholes with a length $\lwormhole$ for
the characteristic dimension of an individual wormhole
and $\lseparation$ for the average separation between
the different wormholes (see text).}
\label{fig:bounds}
\end{figure}

A  concrete example of this particular spacetime model with permanent
wormholes would then have
\begin{equation}  
\lwormhole \approx 10\times l_{\rm Planck}\,,\quad
\lseparation \gtrsim 10^{8} \times l_{\rm Planck} \,,
\end{equation}
in order to be consistent with the bounds (\ref{lfoambound}) and
(\ref{gamma1bound}) for $A=\alpha$. More generally,
Fig.~\ref{fig:bounds} shows which combinations of values
of $\lwormhole$ and $\lseparation$ are allowed or excluded,
assuming $A=\alpha$.
For $\lwormhole \lesssim 1.3 \times 10^{-25}\, {\rm cm}$
(or  $\lseparation \lesssim 1.5\times 10^{-19} \, {\rm cm}$)
the bound (\ref{lfoamgamma1bounds}b) is seen to be the stronger one
and for the other case the bound (\ref{lfoamgamma1bounds}a).
Without further input, we cannot say anything
about $\lwormhole$ and $\lseparation$ individually.

\section{Discussion}

Using experimental bounds \cite{GagnonMoore}
on possible Lorentz-violating modifications of the photon
dispersion law from the absence of Cherenkov-like processes for
high-energy cosmic rays, we have obtained bounds on the length scales
of a photon-propagation model (\ref{Photonmodel}) with time-independent
random background field, which could result from a
static, multiply connected spacetime foam \cite{KlinkhamerRupp}.
Even though the effective length scale $\lgamma$ which enters the
photon dispersion law is constrained to be  below the Planck length
$\lplanck$ for $A=\alpha$,
these bounds do not rigorously exclude a foamlike structure of
spacetime with length scales $\lwormhole$ and $\lseparation$
at or even above the Planck length (see Fig.~\ref{fig:bounds}).

On the other hand, it would perhaps not be unreasonable to
expect \cite{Wheeler} some remnant  ``quantum-gravity'' effect
with \emph{both} length scales $\lwormhole$ and $\lseparation$ of the order
of the Planck
length (\ref{lplanck}), even for a time-independent model with corresponding
preferred frame of reference. But the static wormhole gas
with $\lwormhole \sim \lseparation \sim  \lplanck \approx
10^{-33}\,\,{\rm cm}$ and $A \sim \alpha$ is ruled out
by the bounds (\ref{lfoamgamma1bounds}ab) in terms of
(\ref{lfoamgamma1generaldef}ab); cf. Fig.~\ref{fig:bounds}.
[The crucial assumption here is that the (static) spacetime foam gives rise
to an effective theory (\ref{Photonmodel}) with $g_1(\vec x\,)$
amplitude $A$ of order
$\alpha$. If, for some reason, $A$ would be very much smaller than $\alpha$,
the bounds (\ref{lfoamgamma1bounds}ab) become essentially inoperative.
As mentioned in the previous section,
the preliminary calculations of Ref.~\cite{KlinkhamerRupp}
do suggest $A\sim\alpha$,  but this remains to be confirmed.]

The tentative conclusion
is, therefore, that a preferred-frame graininess of space
with a single length scale $\lplanck$ may be hard to reconcile
with the current experimental bounds from cosmic-ray physics.
Without fine-tuning, such a graininess of
space can also  be expected \cite{Collins-etal} to show up in
``low-energy''  physics (i.e., $\sqrt{s} \ll E_{\rm Planck}
\equiv \sqrt{\hbar\,c^5/G} \approx 1.2 \times 10^{19} \, {\rm GeV}$)
with powers of the coupling constants as the only suppression factor,
an example being the linear term of Eq.~(\ref{dispersionlaw})
with $A^2\,\gamma_1 \sim \alpha^2$.
One possible solution would have gravity as an emergent phenomenon
and the Lorentz-violation scale moved
to trans-Planckian energies \cite{KlinkhamerVolovik}.
But, this is only one out of many suggestions and the puzzle of the apparent
smoothness of space remains unsolved.

\noindent\section*{ACKNOWLEDGMENTS}

FRK thanks G.E. Volovik for interesting discussions and,
also, S. Liberati, V.A. Rubakov, and the other participants of
the Conference on Fundamental Symmetries and Fundamental Constants
(ICTP, Trieste, September 2004).
CR thanks A. Mazumdar for useful comments after a seminar
(NBI, Copenhagen, April 2005).

\end{document}